\documentstyle{article}
\begin{document}
\section{Introduction}
 
The Lattice Boltzmann Method (LBM) was introduced in the late $80$'s
to cope with the two major drawbacks of Lattice Gas Cellular Automata
(LGCA), i.e. statistical noise and exponential complexity of the
updating rule governing the time evolution of the cellular automaton
(Mc Namara 1988, Higuera \& Jimenez 1989, Benzi et al. 1992).
Ever since, the LBM has undergone progressive refinements which have
brought it to the point
where it can compete with most advanced computational fluid
dynamics (CFD) methods for a wide variety of problems, ranging from
fully developed, homogeneous incompressible turbulence, to
low Reynolds number flows in porous media.\\
However, when it comes to complex geometries such as those
commonly encountered in many engineering applications, for instance
internal flows of automotive interest, LBM still lags significantly
behind state-of-the art CFD techniques.
This is due to the inability of LBM to accommodate any sort of
non-uniformity in the spatial distribution  of the mesh grid points.
This limitation is a direct inheritance from LGCA, which are based
upon a set of mono-energetic particles (same speed amplitude for the
various propagation directions) hopping synchronously, in lock-step mode,
from site to site according to the direction of the discrete speeds.
Since the discrete speeds must be the same at any lattice
site, a uniform spatial lattice necessarily results.\\
This is in a blatant contrast with the modern CFD methods
which are generally capable of accomodating fairly complex meshes.
In the attempt to bridge this gap, a coarse-grained extension of the LBM
has been recently introduced (Nannelli 1992).\\
This extension, by borrowing standard ideas from the Finite Volume
method, does provide a significant enhancement of the geometrical
flexibility of LBM, although, for the sake of simplicity,
it was restricted to two dimensional, cartesian non-uniform grids.\\
In this paper, the two dimensional restriction is lifted and a fully
three dimensional coarse-grained LBM is developed.
Before moving on to the details on how this is achieved, 
let us spend some
remarks on a further reason why, we believe, the present study is 
warranted.
\\
It is argued (Boris 1989, Boris et al. 1992) that the increasing availability of parallel
computers is pushing CFD towards a situation of diminishing returns
in terms of trading off computational cost for accuracy.
In other words, the question is whether it is more effective to increase
the grid resolution using a low-order ``lean'' scheme, rather then striving
to save memory using a high-order ``heavy'' scheme.
\\
The LB method is well positioned to attempt a contribution in this direction:
it is a low-order explicit scheme (2nd order in space, first in time)
which performs extremely well on virtually any parallel architecture.
On the other hand, as it stands today, the LB method cannot compete 
with modern CFD methods in situations where non-uniform stretched mesh are
required.
In fact, the gap in number of grid points is simply too huge and no
argument of better parallel efficiency can really compensate for it.
This prompts the need of developing extended LB schemes able to reduce 
the gap, if not close it altogether.
\\
The extension we shall be looking for, should be such as to achieve geometrical
flexibility without compromising the outstanding amenability to parallel
computing to any serious extent.
This paper presents the first exploratory effort in this direction for
flows of engineering interest.
 
\section{Short review of Lattice Boltzmann method}
 
The Lattice Boltzmann Equation reads as follows:
\begin{equation}
f_i ( \vec{x}+\vec{c}_i , t+1 ) - f_i ( \vec{x} , t ) =
\sum_{j=1}^b \Omega_{ij} ( f_j - f_j^{eq} )
\label{eqa4}
\end{equation}
\noindent
where $ f_{i} $ represent the probability of a particle to be moving 
along direction $ \vec{c}_{i} $, 
$\Omega_{ij}$ is the scattering matrix between state i and j
and $ f_j^{eq} $ are the local equilibrium populations, expanded to
second order in the flow speed $ \vec{u} $ to retain convective effects.
Here $ f_{j}^{eq} $ is given by (repeated indices are summed upon):
\begin{equation}
f_{i}^{eq} = \frac{\rho}{b} \left( 1 + \frac{1}{2} c_{i,\alpha}
            u_{\alpha} + 2 Q_{i, \alpha \beta}
            u_{\alpha} u_{\beta} \right)
\;\;\;\;\;\;\;  \alpha, \beta = x,y,z
\label{eqa4_1}
\end{equation}
\noindent
where  $ Q_{i,\alpha \beta} \equiv \  c_{i,\alpha} c_{i,\beta} -
\frac{1}{2} \delta_{\alpha \beta} $, is the projector upon the i-th
speeds, $ \rho \equiv \sum_{i} f_{i} $ is the flow density, and
$ \vec{u} \equiv \sum_{i} f_{i} \vec{c}_{i} / \rho $
is the hydrodynamic velocity.
The discrete speeds $ \vec{c}_{i} $ belong to a four-dimensional 
face centered hypercube (FCHC) defined by 
$ |\vec{c}_{i}| = \sqrt{2} $, and  
$ c_{i,x}+ c_{i,y}+c_{i,z}+c_{i,w} = 2 $ (Frisch et al., 1987).
\\
The equation (\ref{eqa4}) can be regarded as an explicit finite-difference
approximation to a model Boltzmann equation of a BGK type (Qian et al. 1992).
Also one can prove
the existence of a H-theorem which guarantees its numerical stability
in the {\it linear} regime (Mach $ \ll 1$) provided the spectrum
of $\Omega_{ij}$ is confined to the strip $ -2 < \lambda < 0 $ .
As a result, the eigenvalue $ \lambda $ can be tuned to minimize the
viscosity according to the relation
\[
\nu = - \frac{1}{3} ( \frac{1}{2} + \frac{1}{\lambda} )
\]
\noindent
For further details see the recent review by Qian et al., (1995).
\\
The basic merits of LBE are:
\begin{itemize}
\item Flexibility in the choice of the collision rules;
\item Flexibility in the handling of boundary conditions;
\item Ideal amenability to vector/parallel computing;
\end{itemize}
A serious drawback of LBE, as compared to advanced CFD solvers,
relates to the constraint of operating on a uniform, regular mesh.
This limitation is particularly offending for those engineering
applications in which a selective distribution of the spatial grid
points is required in order to cluster the degrees of freedom there
where needed on account of physical and geometrical demands.
\\
The main idea proposed in this paper is to overcome the
limitation of the uniform LB scheme based on the
\underline{two-grid} procedure (Nannelli 1992, Succi 1994)
described in the next section.
 
\section{LB for a non-uniform grid}           \label{n-u_grid}
 
Think of two different lattices, $ L_{f} $ and $ L_{c} $:
$ L_{f} $ is a fine-grained uniform lattice corresponding to
the usual LB scheme, $ L_{c} $, instead, is a non-uniform coarser
lattice whose cells typically contain several nodes of $ L_{f} $.
The idea is to take the differential form of LB Dynamics:
\begin{equation}  \label{eq1}
  \partial_{t} f_{i} + \vec{c}_{i} \cdot \vec{\nabla} f_{i} =
  \sum_{j=1}^{b} \Omega_{i j} (f_{j} - f_{j}^{eq}) \equiv \omega_{i}
\end{equation}
\noindent
and apply a finite-volume procedure based upon integration
of eq. (\ref{eq1}) on each cell of the coarse grid $ L_{c} $.
By straightforward use of Gauss theorem, we obtain
\begin{equation}  \label{eq2}
  \frac{d F_{i}}{d t} + \Phi_{i} = \Omega_{i}  \;\;\;\;\;, i=1,b
\end{equation}
\noindent
where
\begin{eqnarray}  \label{eq3}
   F_{i}      & = & \frac{1}{V_{c}} \int_{C} f_{i} d^{3}x \\
   \Phi_{i}   & = & \frac{1}{V_{c}} \int_{\partial{C}} f_{i}
                   (\vec{c}_{i} \cdot \hat{n}) d^{2}x \nonumber\\
   \Omega_{i} & = & \frac{1}{V_{c}} \int_{C}
                    \omega_{i}  d^{3}x  \nonumber
\end{eqnarray}
\noindent
Here $ F_{i} $ is the mean population of the macrocell $ C $,
$ \Phi_{i} $ the corresponding flux across the boundaries of
$ C $, and $ \Omega_{i} $ is the rate of change of $ F_{i} $
due to collisions occurring within the cell $ C $.
\\
The expression (\ref{eq2}) represents a set of $ N_{c} $ ordinary
differential equations for the unknowns $ F_{i} $, $ N_{c} $
being the number of cells of the coarse grid $ L_{c} $.
To close this system, we need to express the surface
fluxes $ \Phi_{i} $ in terms of the cell-values $ F_{i} $.
This calls for an appropriate interpolation procedure mapping
the fine-grain distribution $ f_{i} $ onto the coarse-grain
distribution $ F_{i} $.
Note that while  $ f_{i} $ is defined on the \underline{nodes}
of $ L_{f} $, $ F_{i} $ is located on the \underline{centers}
of the cells of $ L_{c} $.
In a formal sense, we can write:
\begin{equation}  \label{eq4}
   \hat{R} : F_{i} \rightarrow f_{i}
\end{equation}
\noindent
where the reconstruction operator  $ \: \hat{R} $ is the
(pseudo-)inverse of the averaging operator $ \hat{A} $:
\begin{equation}  \label{eq5}
   \hat{A} : f_{i} \rightarrow F_{i}
\end{equation}
\noindent
operationally defined by eq. (\ref{eq3}). \\
Two reconstruction operators have been
considered: a piece-wise constant (PWC) and a piece-wise
linear (PWL). For the sake of simplicity
hereafter we shall restrict our attention to the case of a
cartesian non-uniform grid stretched along the z-direction:
\begin{equation}  \label{eq6}
   \Delta x = b_{x} \cdot a,\hspace{1 cm}  
   \Delta y = b_{y} \cdot a,\hspace{1 cm}  
   \Delta z = h(z)  \cdot a
\end{equation}
\noindent
where $ a $ is the spacing of the uniform fine-grained grid.
\\
This extends previous non-uniform LB schemes in three respects:
first, the scheme is three dimensional; second, the stretching
factors $ b_{x}, b_{y}, h(z) $ need not be integers; third,
the stretching factor along $ z $ need not be a constant.
\\
By using piece-wise constant interpolation for the collision operator
(local in space) and piecewise linear for the streaming operator (first-order),
we arrive at the following coarse-grained LB equation 
(\underline{F}inite \underline{V}olume \underline{LBE}, FVLBE for short):
\begin{equation}  \label{eq17}
\hat{F}_{i} = \sum^{N_{i}}_{\nu = 0} C_{i}^{\nu} F_{i}^{\nu} +
              \sum_{j}^{b} \Omega_{ij}
              ( F_{j} - F_{j}^{eq} )
\end{equation}
\noindent
where
$ \hat{F}_{i} $ is the mean popolation at time $ t + \Delta t$,
the index $ \nu $ denotes the macrocell involved in the piecewice linear
interpolation (i.e. $ F_{i}^{0} \equiv F_{i} $), and $ C_{i}^{\nu} $ 
are the coarse-grained streaming coefficents.
The scattering matrix $\Omega_{ij}$ is the same as 
given in eq. \ref{eqa4}, and 
$ F_{i}^{eq} $ is the same as eq. \ref{eqa4_1}.
The streaming coefficents $ C_{i}^{\nu} $ are given in Appendix for
propagation direction $ \vec{c}_i = (0,0,1) $ and
$ \vec{c}_{i} = (1,0,1) $.
\\
Consistently with the intents stated in the introduction, the eq.\ref{eq17}
achieves geometrical flexibility at a minimal price in terms of aptness
to parallel computing.
Geometrical flexibility is in charge of coarse-grained streaming coefficents 
$ C_{i}^{\nu} $, while almost ideal amenability to parallel computing is 
preserved because within our piecewise interpolation, the coarse-grained
collision is still completely local in space.

\section{Turbulent Channel Flow simulations}  \label{c-flow}
 
The FVLBE scheme described in the previous section has been
tested for the case of three dimensional turbulent channel flow.
This application is especially suited to our purpose, first 
because it represents the simple instance of a geometry calling 
for a non-uniform mesh, and second in view of the wide body of avalaible
literature.
Two series of simulations have been performed: low resolution
($32 \times 32 \times 100 $) and moderate resolution
($ 64 \times 64 \times 128 $).
Let us begin by describing the former.
Three series of low resolution runs have been performed with
varying factors $ b_{x}, b_{y} $ (see table $1$).
The mesh-distribution along $ z $ is given by the following
$ 1-2-1 $ law,
\[
\Delta z(k) =  \left\{
\begin{array}{ll}
 1 & \mbox{ for $ 1 \le k \le 25 $ }  \\
 2 & \mbox{ for $ 26 \le k \le 75 $}  \\
 1 & \mbox{ for $ 76 \le k \le 100 $}
\end{array}   \right.
\]
\noindent
corresponding to a channel height $ H = 150 $.
The initial condition corresponds to a Poiseuille flow along
$ x $ with average speed $ U_{0} = 0.35 $, perturbed with a
multiperiodic divergence free velocity field.
The molecular viscosity is $ \nu_{0} = 0.005 $, corresponding to a nominal
Reynolds number $ Re_{0} = U_{0} H / 2 \times \nu_{0} \sim 6000 $.
The first outcome of the numerical experiments is that fully developed
turbulence is supported only within a restricted time window, lasting up
to 5000, 15000, 70000 time step for L3, L2, L1 respectively.
As a result only L1 lends itself to a (partial) statistical
analysis of the turbulent field.
The analysis proceeds as follows.
Based on consolidated wisdom, the velocity field in a turbulent
channel flow is expressed as follows:
\begin{equation}  \label{eq18}
\bar{u}_{x}(z) = \left\{
\begin{array}{ll}
 z \frac{v_{*}^{2}}{\nu}  & \mbox{, $ 0 < z <  \delta $} \\
 \frac{v_{*}}{\chi} \log \left(\frac{v_{*} z }{ \nu } \right) + v_{*} d
  & \mbox{, $ z >  \delta $ }
\end{array}   \right.
\end{equation}
\noindent
where $ \chi = 0.4 $ is the Von Karman constant, $ v_{*} $ a typical
turbulent velocity, and $ d $ is a calibration constant
($ d = 5.5 \pm 0.5 $) (Landau).
Here $ \delta = \nu / v_{*} $ is the width of the
``viscous sublayer'' while for $ \delta \le z $ we have the
``inertial sublayer''.
The average velocity profiles drawn from the numerical simulation
are checked against eq. (\ref{eq18}) to produce best fit values of
$ \nu^{n}, v_{*}^{n}, d^{n} $ where the superscript $ n $
denotes ``numerical simulation''.
The consistency check consists of comparing $ \nu^{n} $ with
the input laminar viscosity $ \nu_{0} $, $ v_{*} $ with the
value provided by the wall stress tensor:
$ v_{*}^{2} = < u_{x}(0) u_{z}(0) > $, and finally $ d^{n} $
with the existing literature, i.e. $ d = 5.5 \pm 0.5 $.
The actual values of $ \nu^{n}, v_{*}^{n}, d^{n} $ are derived
from the slope of the linear $ \bar{u}_x $ versus $ z $ plot
($ v_{*}^{2}/ \nu $), the slope of the $ \bar{u}_{x} $ versus $ log(z) $
plot ( $ v_{*} / \chi $ ) and the value of $ log (\bar{u}_{x}) $
at $ z = 1 $ ($ v_{*} / \chi \cdot log(v_{*} / \nu) + d v_{*}$).
In order to assess the grid independence of the numerical results,
three mesh-distributions along $ z $ have been examined:
\\
Lattice ``A''
\[
\Delta z(k) =  \left\{
\begin{array}{ll}
 1 & \mbox{ for $ 1 \le k \le 24 $ }  \\
 2 & \mbox{ for $ 25 \le k \le 75 $}  \\
 1 & \mbox{ for $ 76 \le k \le 100 $}
\end{array}   \right.
\]
\\
\\
Lattice ``B''
\[
\Delta z(k) =  \left\{
\begin{array}{ll}
 1 & \mbox{ for $ 1 \le k \le 14 $ }  \\
 from \;\; 1.05 \;\; to \;\; 1.95  & \mbox{ for $ 15 \le k \le 34 $}  \\
 2 & \mbox{ for $ 35 \le k \le 65 $}   \\
 from \;\; 1.9 \;\; to \;\; 1.1  & \mbox{ for $ 66 \le k \le 85 $}  \\
 1 & \mbox{ for $ 86 \le k \le 100 $ }  \\
\end{array}   \right.
\]
\\
\\
Lattice ``C''
\[
\Delta z(k) =  \left\{
\begin{array}{ll}
 1 & \mbox{ for $ 1  \le k \le 37 $}  \\
 2 & \mbox{ for $ k =  38 $}  \\
 3 & \mbox{ for $ 39 \le k \le 62 $}  \\
 2 & \mbox{ for $ k =  63 $}  \\
 1 & \mbox{ for $ 64 \le k \le 100 $}
\end{array}   \right.
\]
\noindent
The time averaged velocity profiles $ \bar{u}_{x}(z)$ are shown
in fig.\ref{figA}, fig.\ref{figAA}, fig.\ref{figAAA}, while
corresponding best-fit values are reported in table $2$.
The dashed line in fig. \ref{figA}, corresponds to the analytical
profiles, eq. (\ref{eq18}), with the min \& max values drawn from the
numerical experiment.
These min \& max are obtained by interpolating the numerical data
with a family of straight lines, and then taking the min \& max
slopes within this family.
The reason for dealing with a family of straight lines instead of
just one, is that there doesn't appear to be a unique choice for the
set of numerical data to be included in the best fit procedure.
Consequently, by reporting both min and max we intend to provide a measure
of the statistical scatter.
From these figures we see that the grids ``A'' and ``B'' are the best
performers, while grid ``C'' is pretty out of target.
No consistency check for $ v_{*} $ is available for these simulations
because the turbulence window is too narrow to allow the collection
of a significant statistical sample for the wall stress tensor.
In summary, grid A provides similar results as grid B, although
slightly better in terms of effective viscosity.
In either cases, the measured viscosity is more than twice
the laminar one $ \nu_{0} $.
This is the result of the non-uniform mesh which introduces
sharp localized peaks of artificial viscosity in the vicinity
of mesh size discontinuities (z= 32 in our case).
This effect has been found to fade away as the grid resolution
is increased (Amati 1994, Succi 1995).
In consequence, moderate resolution runs have been performed using
grid A.
 
\subsection{Moderate Resolution simulations} \label{simulations}
 
These simulations have been performed on a $ 64 \times 64 \times 128 $
grid with scaling factors $ b_{x} = 15, b_{y} = 8 $.
Mesh points along $ z $ have been distributed according to
the following $ 1-2-1 $ law:
\[
\Delta z(k) =  \left\{
\begin{array}{ll}
 1 & \mbox{ for $ 1 \le k \le 32 $ }  \\
 2 & \mbox{ for $ 33 \le k \le 96 $}  \\
 1 & \mbox{ for $ 97 \le k \le 128 $}
\end{array}   \right.
\]
\noindent
The result is a physical channel of height $ H = 192 $, length
$ L_{x} = 960 $, and width $ L_{y} = 512 $, i.e. pretty close
to the one examined by Moin and coworkers (Moin \& Kim 1980, Moin \& Kim 1982,
Rogers \& Moin 1987, Kim et al. 1987, Jimenez \& Moin 1991).
The main outcome of these simulations is that turbulence is supported for
the entire life span of the simulation, that is $2.4 \times 10^{5}$ time steps,
corresponding to about $ 90 $ transit time units $ L_{x} / U_{0} $.
This is due to the fact that the channel length is now able
to support streamwise rolls feeding cross-channel turbulence.
These samples have been collected every $ 53 $ steps in the interval
$[ 10^{5}, 2.4 \times 10^{5} ]$, thus  
yielding about $ 2600 $ profiles
for statistical analysis.
The results are shown in fig. \ref{fig1} and fig. \ref{fig2}.
The numerical best-fit values deduced from the analysis are as follows:
\begin{equation}  \label{eq27}
\nu^{n}   =  0.013 \pm 0.002, \;\;\;
v_{*}^{n} =  0.013 \pm 0.001, \;\;\;
d^{n}     =  6.5   \pm 0.7
\end{equation}
\noindent
As a first remark, we see that $ v_{*}^{n} $ is quite close to the
values provided by low resolution simulations.
Also, we note that $ d^{n} $ is within the error bars
provided by the literature although somewhat ($ 10-20 \% $) too large.
Finally, since turbulence is sustained for a significant time-span,
wall stress-tensor statistics is also available.
This yields:
\begin{equation}  \label{eq28}
v_{*} \equiv \sqrt{< u_{x} u_{z} >} |_{z = 0}   \sim 0.012
\end{equation}
\noindent
in a pretty good match with the values deduced from the velocity
profiles.\\
For the sake of a better comparison with the existing literature, rescaled
data $u^{+} = u/v_{*}$ and $\tau^{+} = <uv>/(v_{*})^{2}$ are reported 
as a function of dimensionless units 
($ z^{+} = z / \delta $ and $(2 z/H -1)$).
These results are presented in fig. \ref{fig1} and \ref{fig2}  
in which the same quantities
pertaining to other numerical and experimental results 
are also reported.
From figure \ref{fig1} we see that our mean flow compares rather well
with existing data, although the overestimation of
$d$ is clearly visible.
\\
This points to a lack of resolution which prevents
our simulation to attain sufficiently high Reynolds numbers.
Note in fact that the thickness of the boundary
layer $\delta = \nu/ v_{*}$ is in our simulation just one lattice spacing wide.
\\
Such a consideration is indeed corroborated by the results
shown in fig. \ref{fig2}.
From this figure we see that while the stress tensor is correctly captured 
in the central region of the channel (whence the possibility to
obtain a correct estimate of $v_*$), the wall turbulence is definitely
too low as compared with literature data.
\\
These moderate resolution runs suggest that the FVLBE scheme provides
results within the errors bars of current CFD methods
even though a better control of numerical diffusion is needed to make
it more competitive.
At this stage, it is therefore of interest to spend some comments
on the issue of numerical efficiency.
As pointed out in the introduction, FVLBE has been generated
in order to extend the range of applicability of the Lattice
Boltzmann method to non-uniform geometries
while keeping optimal amenability to parallel computing.
This is achieved at the expense of an increased compute density,
i.e. floating point operation per grid point,
because, at variance with LBE, the propagation step involves more
than just a two-point stencil.
The idea is to offset this excess of computation per node
by a substantial reduction of the number of grid points to be used
in the simulation, which is made possible by the capability
to compress/rarefy the spatial grid distribution.
For the case in point, substantial grid savings
should be planned along the streamwise ($ x $) and spanwise
($ y $) direction where relatively long-wavelength structures
are expected to arise as compared to cross-flow ($ z $) eddies.
For the present $ 64 \times 64 \times 128 $ simulation, each time step
takes about $ 5 $ second CPU time on a IBM RISC/6000 mod.580.
This corresponds to about $ 10 \mu s $ per grid point per step,
to be compared with roughly $ 3 \mu s $ taken by a uniform LBE.
These figures reflect approximately the increase in the number of
floating point operation per grid point:
about 1000 for FVLBE and 500 for LBE.
This factor two is largely overcompensated by the much larger size
accessible to FVLBE, i.e. $ L_{x} = 960, L_{y} = 512, H = 192 $,
corresponding to a gain factor of
\begin{equation}  \label{eq29}
\frac{512}{64} \cdot \frac{960}{64} \cdot \frac{192}{128} = 180
\end{equation}
\noindent
i.e. about two orders of magnitude.
Indeed the channel flow simulation presented in this paper would be
simply \underline{unfeasible} with a plain LBE scheme,
for the latter would take about $ 30000 $ CPU seconds per time step,
and about $ 150 $ Gbyte of storage!
\\
To date, the largest channel flow simulation we have been also able 
to perform with a \underline{uniform} LB scheme is a 
$ 432 \times 144 \times 288 $ (1.7 GB) corresponding to $ Re  \simeq 3000 $,
using the  512 processor Quadrics Machine (Bartoloni et al., 1993)
Although the parallel performance is exceedingly encouraging
(parallel efficiency $ 54 \; vs. \; 64 $, 
as can be seen in fig.\ref{figAAAA}, Amati et al., 1996).
It is clear that parallel computing alone cannot make up for 
the overdemand of computational resources raised by uniform LB scheme.
\\
Our code is almost a factor ten faster than modern semi-implicit CFD methods
(compare our 5 s/step on a $ 64 \times 64 \times 128 $ with a 25 s/step
on a $ 32 \times 64 \times 97 $, see Orlandi, 1995) but the quality 
of the results is correspondingly less satisfatory.
Both gaps are likely to close up once better interpolators are in place.
At this stage, only aptness to parallel computing will make the difference.

\section{Conclusions}                         \label{conclusion}
 
By borrowing standard techniques from the finite volume method,
a low order coarse-grain three dimensional LB scheme has been
developed.
\\
This scheme basically preserves the outstanding amenability to
parallel computing of the uniform LB scheme,  while giving access to 
a much larger Reynolds number class of flows.
Actual numerical simulation do, however, reveal that the large-scales  
(the resolved ones) display less turbulent activity than expected
on the basis of the nominal Reynolds number.
This means that, while marking a singificant stride forward with respect
to the uniform scheme, the present coarse-grained LB still lags 
behind state-of-art CFD methods.
\\
A plausible explanation is that our low order interpolator
(piecewise constant for collision operator, and piecewise linear 
for the streaming operator) does achieve locality (hence aptness to
parallel computing) much at the expense of accuracy.
Future work shall then focus on the development of better interpolation
schemes, possibly in the spirit of  Monotone Interpolated 
Large Eddies Simulation (MILES).
\\
In principle there is no reason why the basic advantages of LBM,
i.e. handy treatment of complex boundary conditions and outstanding
amenability to parallel computing, should not carry over into FVLBE.
Should this be the case, FVLBE may represent a fairly competitive
tool for the numerical investigation of inhomogenous turbulent
flows on highly parallel machines.
 
\section{Acknowledgements}
 
The authors are indebted to Prof. V. Yakhot, Y. H. Qian and S. Orszag
for illuminating discussions.
\\
Prof. P. Orlandi is kindly acknowledged for providing literature data
reported in Figs. \ref{fig1}, \ref{fig2}.
G.A. would to acknowledge the hospitality of IBM ECSEC where numerical
work has been performed.
\\
Parallel simulations were performed on the ENEA Quadrics machine.

%appendice: coefficienti per l'operatore di propagazione
\appendix
 
\section{Appendix} \label{app}
 
In this Appendix we report the explicit expressions of some representative
streaming coefficients.
All quantities are measured in units of the fine-grain uniform
lattice, i.e. $ a = 1 $.  
Along the direction $ \vec{c}_{i} = (0,0,1)$ the evolution of
$ F $ is given by:
\begin{eqnarray*}
\hat{F} & = &  \alpha F + \beta F^{I} + \gamma F^{II}
\end{eqnarray*}
\noindent
where $ F^{I} $ and $ F^{II} $ are the population 
at the nearest and next-to-nearest left neighbors
and $ \alpha $, $\beta $ and $ \gamma $ are 
the streaming coefficients. Their expression is: 
\begin{equation}   \label{a6}
 \alpha(k) = \left[ 1 -  \frac{1}{ h(k)} - \frac{1}{ h(k)}
              \frac{2 ( z^{+}(k)-z(k) )}
                      {\ h(k)+ h(k-1)} \right]
\end{equation}

\begin{equation}   \label{b6}
\beta(k) = \frac{1}{ h(k)} \left[ 1 + \frac{2 (z^{+}(k)-z(k) )}
{h(k)+h(k-1)} + \frac{2(z^{+}(k-1)-z(k-1))}{h(k-1)+h(k-2)}\right]
\end{equation}

\begin{equation}    \label{c6}
\gamma(k) =  -\frac{2(z^{+}(k-1)-z(k-1))}{ h(k-1)+ h(k-2)}
              \frac{1}{h(k)}
\end{equation}
\noindent
here the coefficients depends on the index $ k $,
$ z(k) $ is the z-coordinate of the center of the macrocell, and
$  z^{+}(k) \equiv z(k) + \frac{h(k)-1}{2} $.
\\
For diagonal propagation, like $ \vec{c}=(1,0,1) $, the
coefficients are:
\begin{equation}
  \hat{F}_{i} = \alpha_{i} F_{i} + \beta_{i} F_{i}^{I}
               + \gamma_{i} F_{i}^{II} + \delta_{i} F_{i}^{III}
               + \epsilon_{i} F_{i}^{IV} + \eta_{i} F_{i}^{V}
\end{equation}
  
\noindent
\begin{eqnarray}
 \alpha_{i} & = &  \left[  1  -  
           \frac{( b_{x} + h(k) -1)}{ b_{x} h(k) } \right] +     
           \frac{b_{x}-\frac{1}{2}}{b_{x} h(k)}
   \left[  \frac{2 ( z^{-}(k)-z(k) )}{h(k) + h(k-1)} \right]\\
 \nonumber \\
  &   &  \mbox{} - \frac{ h(k)-\frac{1}{2}}{b_{x} h(k)}
          \left[  \frac{ ( x^{+}(k)-x(k) )}{b_{x}} \right]  \nonumber
\end{eqnarray}
  
\begin{eqnarray}
 \beta_{i} & = &
  \frac{  b_{x}  - \frac{1}{2}}  { b_{x} h(k) }
         + \frac{ b_{x}-\frac{1}{2}}{b_{x} h(k)}
                \left[  \frac{2 ( z^{-}(k)-z(k) )}
                 { h(k) + h(k-1)} \right] + \\
 \nonumber \\
  &   &  \mbox{} + \frac{ b_{x}-\frac{1}{2}}{b_{x} h(k)}
                \left[  \frac{2 ( z^{-}(k+1)-z(k+1) )}
                 { h(k+1) + h(k)} \right]  \nonumber
\end{eqnarray}
  
\begin{equation}
 \gamma_{i}  =  - \frac{ b_{x}-\frac{1}{2}}{b_{x} h(k)}
             \left[  \frac{2 ( z^{-}(k+1)-z(k+1) )}
                          { h(k+1) + h(k)} \right]\\
\end{equation}
  
\begin{eqnarray}
 \delta_{i}  & = &  \frac{ h(k) - \frac{1}{2}} { b_{x} h(k) }
         + \frac{ h(k)-\frac{1}{2}}{b_{x} h(k)}
          \left[  \frac{ ( x^{+}(i)-x(i) )}{b_{x}} \right] + \\
\nonumber \\
  &   &  \mbox{} + \frac{ h(k)-\frac{1}{2}}{b_{x} h(k)}
       \left[ \frac{ ( x^{+}(i+1)-x(i+1) )}{b_{x}} \right] \nonumber
\end{eqnarray}
  
\begin{equation}
 \epsilon_{i} =   - \frac{ h(k)-\frac{1}{2}}{b_{x} h(k)}
       \left[ \frac{ ( x^{+}(i+1)-x(i+1) )}{b_{x}} \right]
\end{equation}
  
\begin{equation}
\eta_{i}  =  \frac{1}{ b_{x} h(k) }
\end{equation}
% 
% THAT'S ALL FOLKS
 
%
% bibiliografia
%

\end{document}